\input epsf							
\newfam\scrfam
\batchmode\font\tenscr=rsfs10 \errorstopmode
\ifx\tenscr\nullfont
        \message{rsfs script font not available. Replacing with calligraphic.}
        \def\scr{\cal}
\else   
        \font\sevenscr=rsfs7
        \font\fivescr=rsfs5
        \skewchar\tenscr='177 \skewchar\sevenscr='177 \skewchar\fivescr='177
        \textfont\scrfam=\tenscr \scriptfont\scrfam=\sevenscr
        \scriptscriptfont\scrfam=\fivescr
        \def\scr{\fam\scrfam}
        \def\cal{\scr}
\fi
\newfam\msbfam
\batchmode\font\twelvemsb=msbm10 scaled\magstep1 \errorstopmode
\ifx\twelvemsb\nullfont\def\Bbb{\bf}

	\message{Blackboard bold not available. Replacing with boldface.}
\else   \catcode`\@=11
        \font\tenmsb=msbm10 \font\sevenmsb=msbm7 \font\fivemsb=msbm5
        \textfont\msbfam=\tenmsb
        \scriptfont\msbfam=\sevenmsb \scriptscriptfont\msbfam=\fivemsb
        \def\Bbb{\relax\expandafter\Bbb@}
        \def\Bbb@#1{{\Bbb@@{#1}}}
        \def\Bbb@@#1{\fam\msbfam\relax#1}
        \catcode`\@=\active

\fi
        \font\eightrm=cmr8              \def\xrm{\eightrm}
        \font\eightbf=cmbx8             \def\xbf{\eightbf}
        \font\eightit=cmti10 at 8pt     \def\xit{\eightit}
        \font\eighttt=cmtt8             \def\xtt{\eighttt}
        \font\eightcp=cmcsc8
        \font\eighti=cmmi8              \def\xold{\eighti}
        \font\eightib=cmmib8             \def\xbold{\eightib}
        \font\teni=cmmi10               \def\old{\teni}
        \font\tencp=cmcsc10
        \font\tentt=cmtt10
        
        \font\twelvecp=cmcsc10 scaled\magstep1

	 at10pt	
	\font\twelvehelvbold=phvb at12pt
	 at14pt
	\font\sixteenhelvbold=phvb at16pt

\def\noblackbox{\overfullrule=0pt}
\noblackbox

\newtoks\headtext
\headline={\ifnum\pageno=1\hfill\else
	\ifodd\pageno{\eightcp\the\headtext}{ }\dotfill{ }{\old\folio}
	\else{\old\folio}{ }\dotfill{ }{\eightcp\the\headtext}\fi
	\fi}
\def\makeheadline{\vbox to 0pt{\vss\noindent\the\headline\break
\hbox to\hsize{\hfill}}
        \vskip2\baselineskip}
\newcount\infootnote
\infootnote=0
\def\foot#1#2{\infootnote=1
\footnote{$\,{}^{#1}$}{\vtop{\baselineskip=.75\baselineskip
\advance\hsize by -\parindent\noindent{\xrm #2}}}\infootnote=0$\,$}
\newcount\refcount
\refcount=1
\newwrite\refwrite
\def\oldsize{\ifnum\infootnote=1\xold\else\old\fi}
\def\ref#1#2{
	\def#1{{{\oldsize\the\refcount}}\ifnum\the\refcount=1\immediate\openout\refwrite=\jobname.refs\fi\immediate\write\refwrite{\item{[{\xold\the\refcount}]} 
	#2\hfill\par\vskip-2pt}\xdef#1{{\noexpand\oldsize\the\refcount}}\global\advance\refcount by 1}
	}
\def\refout{\catcode`\@=11
        \xrm\immediate\closeout\refwrite
        \vskip2\baselineskip
        {\noindent\twelvecp References}\hfill\vskip\baselineskip
        \baselineskip=.75\baselineskip
        \input\jobname.refs
        \baselineskip=4\baselineskip \divide\baselineskip by 3
        \catcode`\@=\active\rm}

\def\hepth#1{\href{http://xxx.lanl.gov/abs/hep-th/#1}{{\xtt hep-th/#1}}}
\def\jhep#1#2#3#4{\href{http://jhep.sissa.it/stdsearch?paper=#2\%28#3\%29#4}{J. High Energy Phys. {\xbold #1#2} ({\xold#3}) {\xold#4}}}
\def\JHEP{\jhep}
\def\PLB#1#2#3{Phys. Lett. {\xbf B}{\xbold#1} ({\xold#2}) {\xold#3}}
\def\NPB#1#2#3{Nucl. Phys. {\xbf B}{\xbold#1} ({\xold#2}) {\xold#3}}
\def\PRD#1#2#3{Phys. Rev. {\xbf D}{\xbold#1} ({\xold#2}) {\xold#3}}

\def\IJMPA#1#2#3{Int. J. Mod. Phys. {\xbf A}{\xbold#1} ({\xold#2}) {\xold#3}}

\def\CQG#1#2#3{Class. Quantum Grav. {\xbold#1} ({\xold#2}) {\xold#3}}
\newcount\sectioncount
\sectioncount=0
\def\section#1#2{\global\eqcount=0
	\global\subsectioncount=0
        \global\advance\sectioncount by 1
	\ifnum\sectioncount>1
	        \vskip2\baselineskip
	\fi
	\noindent
        \line{\twelvecp\the\sectioncount. #2\hfill}
		\vskip.8\baselineskip\noindent
        \xdef#1{{\old\the\sectioncount}}}
\newcount\subsectioncount
\def\subsection#1#2{\global\advance\subsectioncount by 1
	\vskip.8\baselineskip\noindent
	\line{\tencp\the\sectioncount.\the\subsectioncount. #2\hfill}
	\vskip.5\baselineskip\noindent
	\xdef#1{{\old\the\sectioncount}.{\old\the\subsectioncount}}}
\newcount\appendixcount
\appendixcount=0
\def\appendix#1{\global\eqcount=0
        \global\advance\appendixcount by 1
        \vskip2\baselineskip\noindent
        \ifnum\the\appendixcount=1
        \hbox{\twelvecp Appendix A: #1\hfill}\vskip\baselineskip\noindent\fi
    \ifnum\the\appendixcount=2
        \hbox{\twelvecp Appendix B: #1\hfill}\vskip\baselineskip\noindent\fi
    \ifnum\the\appendixcount=3
        \hbox{\twelvecp Appendix C: #1\hfill}\vskip\baselineskip\noindent\fi}
\def\acknowledgements{\vskip2\baselineskip\noindent
        \underbar{\it Acknowledgements:}\ }
\newcount\eqcount
\eqcount=0
\def\Eqn#1{\global\advance\eqcount by 1
\ifnum\the\sectioncount=0
	\xdef#1{{\old\the\eqcount}}
	\eqno({\oldstyle\the\eqcount})
\else
        \xdef#1{{\old\the\sectioncount}.{\old\the\eqcount}}
        \ifnum\the\appendixcount=0
                \eqno({\oldstyle\the\sectioncount}.{\oldstyle\the\eqcount})\fi
        \ifnum\the\appendixcount=1
                \eqno({\oldstyle A}.{\oldstyle\the\eqcount})\fi
        \ifnum\the\appendixcount=2
                \eqno({\oldstyle B}.{\oldstyle\the\eqcount})\fi
        \ifnum\the\appendixcount=3
                \eqno({\oldstyle C}.{\oldstyle\the\eqcount})\fi
\fi}
\def\eqn{\global\advance\eqcount by 1
\ifnum\the\sectioncount=0
	\eqno({\oldstyle\the\eqcount})
\else
        \ifnum\the\appendixcount=0
                \eqno({\oldstyle\the\sectioncount}.{\oldstyle\the\eqcount})\fi
        \ifnum\the\appendixcount=1
                \eqno({\oldstyle A}.{\oldstyle\the\eqcount})\fi
        \ifnum\the\appendixcount=2
                \eqno({\oldstyle B}.{\oldstyle\the\eqcount})\fi
        \ifnum\the\appendixcount=3
                \eqno({\oldstyle C}.{\oldstyle\the\eqcount})\fi
\fi}
\def\multi{\global\advance\eqcount by 1}
\def\multieq#1#2{\xdef#1{{\old\the\eqcount#2}}
        \eqno{({\oldstyle\the\eqcount#2})}}
\newtoks\url
\def\Href#1#2{\catcode`\#=12\url={#1}\catcode`\#=\active#2}
\def\href#1#2{{#2}}

\parskip=3.5pt plus .3pt minus .3pt
\baselineskip=14pt plus .1pt minus .05pt
\lineskip=.5pt plus .05pt minus .05pt
\lineskiplimit=.5pt
\abovedisplayskip=18pt plus 4pt minus 2pt
\belowdisplayskip=\abovedisplayskip
\hsize=14cm
\vsize=19cm
\hoffset=1.5cm
\voffset=1.8cm
\frenchspacing
\footline={}
\def\ss{\scriptstyle}

\def\*{\partial}
\def\punkt{\,\,.}
\def\komma{\,\,,}

\def\={\!=\!}
\def\small#1{{\hbox{$#1$}}}
\def\half{\small{1\over2}}
\def\fraction#1{\small{1\over#1}}
\def\fr{\fraction}
\def\Fraction#1#2{\small{#1\over#2}}
\def\Fr{\Fraction}

\def\eg{{\tenit e.g.}}
\def\ie{{\tenit i.e.}}
\def\etal{{\tenit et al.}}

\def\nlni{\hfill\break}

\def\a{\alpha}
\def\b{\beta}
\def\c{\gamma}
\def\d{\delta}

\def\g{\gamma}
\def\l{\lambda}

\def\G{\Gamma}
\def\L{\Lambda}

\def\H{{\Bbb H}}

\def\Dslash{D\hskip-6.5pt/\hskip1.5pt}

\def\tF{\tilde F}
\def\tJ{\tilde J}

\def\lra{\longrightarrow}
\def\D{\Delta}
\def\Dlra#1{\buildrel\D_{#1}\over\lra}
\def\H{{\cal H}}

\def\k{\kappa}


\headtext={Cederwall, Nilsson, Tsimpis: ``The structure of maximally
supersymmetric$\ldots$''}

\null\vskip-2cm
\line{
\epsfysize=1.7cm
\epsffile{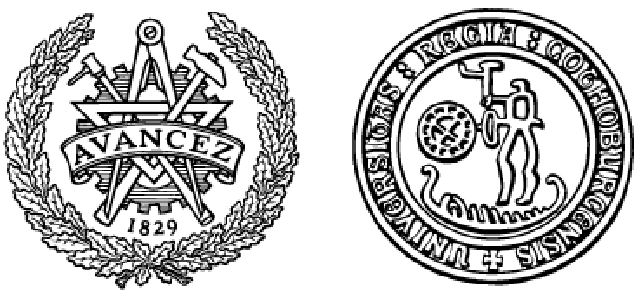}
\hfill}
\vskip-1.7cm
\line{\hfill G\"oteborg ITP preprint}
\line{\hfill\tt hep-th/0102009}
\line{\hfill February {\old2}, {\old2001}}
\line{\hrulefill}

\vfill

\centerline{\sixteenhelvbold The structure of} 
\vskip6pt 
\centerline{\sixteenhelvbold maximally supersymmetric Yang--Mills theory:}
\vskip6pt 
\centerline{\sixteenhelvbold constraining higher-order corrections}

\vskip1.6cm

\centerline{\twelvehelvbold Martin Cederwall,
	 Bengt E.W. Nilsson and Dimitrios Tsimpis}

\vskip.8cm

\centerline{\it Department of Theoretical Physics}
\centerline{\it G\"oteborg University and Chalmers University of Technology }
\centerline{\it SE-412 96 G\"oteborg, Sweden}

\vskip1.6cm

\noindent\underbar{Abstract:} We solve the superspace Bianchi identities 
for ten-dimensional supersymmetric Yang--Mills theory without imposing 
any kind of constraints apart from the standard conventional one.
In this way we obtain a set of algebraic conditions on certain fields 
which in the on-shell theory are constructed as composite ones out of
the physical fields. These conditions must hence be satisfied by any 
kind of theory in ten dimensions invariant under supersymmetry and some, 
abelian or non-abelian, gauge symmetry. 
Deformations of the ordinary SYM theory (as well as the fields) 
are identified as elements
of a certain spinorial cohomology, giving control over field redefinitions
and the distinction between physically relevant higher-order corrections
and those removable by field redefinitions.
The conditions derived severely constrain theories involving $F^2$-level 
terms plus higher-order corrections, as for instance those derived 
from open strings as effective gauge theories on D-branes.

\vfill

\line{\hrulefill}
\catcode`\@=11
\line{\tentt martin.cederwall@fy.chalmers.se\hfill}
\line{\tentt bengt.nilsson@fy.chalmers.se\hfill}
\line{\tentt tsimpis@fy.chalmers.se\hfill}
\catcode`\@=\active

\eject

\ref\NilssonSYM{B.E.W.~Nilsson, 
\xit ``Off-shell fields for the 10-dimensional supersymmetric 
Yang--Mills theory'', \xrm G\"oteborg-ITP-{\xold81}-{\xold6};
{\xit ``Pure spinors as auxiliary fields in the ten-dimensional 
supersymmetric Yang--Mills theory''},
\CQG3{1986}{{\xrm L}41}.}

\ref\Dp{M. Cederwall, A. von Gussich, B.E.W. Nilsson and A. Westerberg,
{\xit ``The Dirichlet super-three-brane in ten-dimensional type IIB 
supergravity''}
\NPB{490}{1997}{163} [\hepth{9610148}];\nlni
M. Aganagi\'c, C. Popescu, J.H. Schwarz,
{\xit ``D-brane actions with local kappa symmetry''},
\PLB{393}{1997}{311} [\hepth{9610249}];\nlni
M. Cederwall, A. von Gussich, B.E.W. Nilsson, P. Sundell
 and A. Westerberg,
{\xit ``The Dirichlet super-p-branes in ten-dimensional type IIA and IIB 
supergravity''},
\NPB{490}{1997}{179} [\hepth{9611159}];\nlni
E. Bergshoeff and P.K. Townsend, 
{\xit ``Super D-branes''},
\NPB{490}{1997}{145} [\hepth{9611173}].}

\ref\LiE{A.M. Cohen, M. van Leeuwen and B. Lisser, 
LiE v. {\xold2}.{\xold2} ({\xold1998}), 
\nlni http://wallis.univ-poitiers.fr/\~{}maavl/LiE/} 

\ref\ConvConstr{S.J.~Gates, K.S.~Stelle and P.C.~West,
{\xit ``Algebraic origins of superspace constraints in supergravity''},
\NPB{169}{1980}{347}; 
S.J. Gates and W. Siegel, 
{\xit ``Understanding constraints in superspace formulation of supergravity''},
\NPB{163}{1980}{519}.}

\ref\CederwallSYMFFOUR{M.~Cederwall, B.E.W.~Nilsson and D.~Tsimpis,
{\xit in preparation}.}

\ref\BergshoeffFFOUR{E.~Bergshoeff, M.~Rakowski and E.~Sezgin,
{\xit ``Higher derivative super Yang--Mills theories},
\PLB{185}{1987}{371}.}

\ref\TseytlinBIREV{A.A.~Tseytlin, 
{\xit ``Born--Infeld action, supersymmetry and string theory''},
\xrm in the Yuri Golfand memorial volume, ed. M. Shifman,
World Scientific (2000) [\hepth{9908105}]}

\ref\SchwarzSBI{M. Aganagi\'c, C. Popescu, J.H. Schwarz,
{\xit ``Gauge-invariant and gauge fixed D-brane actions''},
\NPB{495}{1997}{145} [\hepth{9612080}].}

\ref\TseytlinSTR{A.A.~Tseytlin, 
{\xit "On the non-abelian generalization of 
Born--Infeld action in string theory"},
\NPB{501}{1997}{41} [\hepth{9701125}].}

\ref\BergshoeffKAPPA{E.A.~Bergshoeff, M.~de~Roo and A.~Sevrin,
{\xit ``On the supersymmetric non-abelian Born--Infeld action''},
\hepth{0011264}.}

\ref\WittenDBRANES{E. Witten, 
{\xit ``Bound states of strings and p-branes''},
\NPB{460}{1996}{335} [\hepth{9510135}].}

\ref\SevrinFSIX{A. Sevrin, J. Troost and W. Troost,
{\xit ``The non-abelian Born--Infeld action at order $\ss F^6$''},
\hepth{0101192}.}

\ref\TaylorNONSTR{A. Hashimoto and W. Taylor IV,
{\xit ``Fluctuation spectra of tilted and intersecting D-branes 
from the Born--Infeld action''},
\NPB{503}{1997}{193} [\hepth{9703217}].}

\ref\CGNN{M. Cederwall, U. Gran, M. Nielsen and B.E.W. Nilsson, 
{\xit ``Manifestly supersymmetric M-theory''}, 
\JHEP{00}{10}{2000}{041} [\hepth{0007035}];
{\xit ``Generalised 11-dimensional supergravity''}, \hepth{0010042}.}

\ref\SuperYM{L. Brink, J.H. Schwarz and J. Scherk, 
{\xit ``Supersymmetric Yang--Mills theories''},
\NPB{121}{1977}{77}.}

\ref\NonlinearSS{P.S. Howe, O. Raetzel and E. Sezgin,
{\xit ``On brane actions and superembeddings''},
\JHEP{98}{08}{1998}{011} [\hepth{9804051}];\nlni
J. Bagger and A. Galperin,
{\xit ``A new Goldstone multiplet for partially broken supersymmetry''},
\PRD{55}{1997}{1091} [\hepth{9608177}];\nlni
T. Adawi, M. Cederwall, U. Gran, M. Holm and B.E.W. Nilsson, 
{\xit ``Superembeddings, non-linear supersymmetry and 5-branes''},
\IJMPA{13}{1998}{4691} [\hepth{9711203}];\nlni 
P. Pasti, D. Sorokin and M. Tonin,
{\xit ``Superembeddings, partial supersymmetry breaking and superbranes''},
\NPB{591}{2000}{109} [\hepth{0007048}].} 

\ref\NilssonSixDSYM{B.E.W. Nilsson, 
{\xit ``Superspace action for a 6-dimensional non-extended supersymmetric
Yang--Mills theory''},
\NPB{174}{1980}{335}.}

\ref\Kitazawa{Y. Kitazawa, 
{\xit ``Effective lagrangian for open superstring from five point function''}, 
\NPB{289}{1987}{599}.}

\section\introduction{Introduction}Gauge and reparametrisation invariant 
theories arise as effective field theories in string theory.
 In the case of open bosonic strings these field theories correspond
to ordinary $F^2$ Yang--Mills theory  
only in the limit of  weak fields. In more general situations, due to the 
appearence of the dimensionful parameter $\a'$,
one typically finds an infinite set of higher-order terms involving
arbitrary powers of the Yang--Mills field strength as well 
as field strengths with
any number of covariant derivatives acting on them. To elucidate the structure
of these effective actions has turned out to be extremely difficult,
and it is only in the abelian case and for constant 
field strengths that we have any kind of understanding of 
the complete structure
of the action. Under precisely these conditions, the open 
bosonic string theory is known to 
generate the Born--Infeld lagrangian 
$$
L=-\sqrt{-det(\eta_{ab}+2\pi\a'F_{ab})}\punkt\Eqn\BIlagr
$$
For a quite comprehensive review of the role of the Born--Infeld action 
in string theory including a large number of references, see 
ref. [\TseytlinBIREV].
The supersymmetric version of this lagrangian is also known and takes the form
$$
L=-\sqrt{-det(\eta_{ab}+2\pi\a'F_{ab}-2(2\pi\a')^2\bar{\l}\G_a\partial_b\l
+(2\pi\a')^4\bar{\l}\G^c\partial_a\l\bar{\l}\G_c\partial_b\l)}
\komma\Eqn\SBIlagr
$$
as can be seen [\SchwarzSBI] by setting $p=9$ in the action obtained
by gauge fixing the reparametrisation invariance and 
$\kappa$-symmetry of the actions for the $D_p$-branes derived in 
[\Dp].

Adding non-abelian Chan--Paton factors to the ends of the open strings 
makes it possible to derive non-abelian versions of the effective actions. 
The importance
of these actions have been highlighted recently in connection with
the solitonic D-brane solutions in string theory. As
explained by Witten in ref. [\WittenDBRANES], 
the non-abelian nature of the gauge theory can in this context 
be understood as arising
from a stack
of branes by  a detailed analysis of all possible
configurations with the two ends of the 
strings ending on different branes in the stack. 

Unfortunately, in the non-abelian case
our knowledge about the effective action 
is at a very rudimentary level and there is as yet no 
situation in which we 
understand its structure
to general order in the fields. In fact there are only partial results up to
order $F^6$ [\TseytlinBIREV,\SevrinFSIX] beyond which
we have not been able to obtain any information. 
In the context of 
the superstring  the situation is basically the same.  However, 
here one may pose interesting questions about
to what extent $\kappa$-symmetry, extra non-linear supersymmetries, and/or 
maximal linear supersymmetry constrain the form of higher-order
corrections to the ordinary $F^2$ super-Yang--Mills theory. In the abelian
case $\kappa$-symmetry [\Dp]
and non-linear supersymmetries [\NonlinearSS] are known to be 
intimately connected to the structure of the Born--Infeld theory. 
Finding non-abelian generalisations of these arguments 
has however turned out to be a problem  related to the
difficult issues that arise when trying to define geometry on 
a non-commutative (curved) spacetime. 
 
It was found some time ago [\BergshoeffFFOUR] that at order $F^4$, assuming
a symmetrised trace over the gauge generators, henceforth denoted $Str$, 
 and starting from the $StrF^4$
term, 
supersymmetry is enough to unambiguously
produce the structure of the non-abelian action.
This is in complete agreement
with Tseytlin's suggestion [\TseytlinSTR] that in the non-abelian 
case it might suffice to consider 
the Born--Infeld lagrangian in (\BIlagr) 
but with all fields in the adjoint of the non-abelian
gauge group and with a totally symmetric trace over group 
generators to eliminate the ordering ambiguities. However, 
there are strong indications that 
the $Str$ prescription does not provide the full structure as obtained 
from string theory at order $F^6$ and higher 
[\TaylorNONSTR,\SevrinFSIX]. Recently Bergshoeff \etal\
[\BergshoeffKAPPA] have found, 
based on an attempt to implement $\kappa$-symmetry in a
non-abelian setting, non-$Str$ terms involving fermions (but no pure $F$-terms)
also at lower order than this.
(The order discussed here is most easily kept track of by giving the fields
$F_{\mu \nu}$, $\lambda_{\alpha}$ and spacetime derivatives canonical 
four-dimensional mass dimensions despite the fact that we will be dealing 
exclusively with ten-dimensional super-Yang--Mills theory. When deriving the 
action from string theory these terms get accompanied by appropriate powers
of $\a'$ which has dimension $(mass)^{-2}$.)

In this paper we will consider ten-dimensional
super-Yang--Mills theory [\SuperYM] and approach the problem of finding the 
constraints  on the possible
higher-order corrections implied by (maximal) supersymmetry 
by embedding the theory in superspace.
The theory is then given in terms of a superfield gauge potential 
$A_A=(A_{a},A_{\a})$, where the indices $a$ and $\a$ refer to the vector and 
spinor representations, respectively. The corresponding field strength
 satisfies ordinary superspace Bianchi identities. 
As we will see in the 
next section, from these identities one can easily derive the field equations
corresponding to the lowest order, \ie, $F^2$, theory, supersymmetry
transformation rules etc. By relaxing the constraint that leads to
these field equations one instead obtains a system of superspace 
equations that any
theory consistent with supersymmetry and gauge invariance must satisfy.
In section {\old3}, we  proceed to solve these superspace equations. The
solution is in the form of a number of algebraic conditions on the various
component fields appearing at different levels in the superfields 
in terms of which the theory is defined. 
Section {\old4} is devoted to a formal development of what we might call
spinorial cohomology, where component fields and deformations are obtained as
elements of cohomology classes under a fermionic exterior derivative.
The understanding obtained elucidates the specific properties enjoyed
by maximally supersymmetric Yang--Mills being an on-shell superspace 
theory and gives a solid underpinning of the approach adopted for finding
physically inequivalent deformations.
Section {\old5} is devoted to some
comments concerning how this setup might be used to obtain information
about possible higher-order 
corrections. Some of these comments are of relevance for D-brane physics 
in string theory and the search for a non-abelian Born--Infeld theory. 
A more comprehensive discussion containing also explicit expressions at the 
next, \ie\ $F^4$, level of corrections 
will be published elsewhere [\CederwallSYMFFOUR].

\section\superspace{Implementation of gauge invariance in superspace}This 
section provides a recapitulation of the superspace techniques
in the context of abelian and non-abelian gauge theories, restricted
to ten dimensions which corresponds to maximal supersymmetry. 
We will first set up our conventions and then proceed to 
derive the equations of motion at order $F^2$
by imposing constraints on the superspace Yang--Mills field stength
[\NilssonSYM] .

If we turn a spacetime vector potential $A_{m}(x)$ into a superfield,
 we obtain the basic object, $A_M(x,\theta)$, from which one may construct
 gauge theories  that are manifestly 
supersymmetric.
Here the index $M$ refers to the pair of curved indices $m,\mu$, 
the first enumerating 
the ten bosonic components and the second the sixteen fermionic ones. 
A superspace one-form is 
constructed by contracting the superspace vector potential with 
$dx^m$ and $d\theta^{\mu}$,
which obey opposite statistics to the coordinates $x^m$ and 
$\theta^{\mu}$. From the abelian
one-form potential $A$, which transforms under gauge transformations 
as $\delta A=d\Phi$ where
$\Phi$ is a scalar superfield and 
$d=dx^m\partial_m+d\theta^{\mu}\partial_{\mu}$,
 we construct the gauge invariant superfield strength
$F=dA$. The corresponding Bianchi identity reads $dF=0$. 
In the non-abelian case
we must of course use the covariant derivative $D=d+A$ instead.

So far no on-shell information has been fed into the equations; 
in fact any supersymmetric
gauge theory must satisfy the superspace Bianchi identity (BI) 
$DF=0$. When analysing the superspace BI one usually considers 
its component equations in tangent space. The reason for this is
 that since the 
tangent space structure group is the ordinary Lorentz group 
Spin(1,9) and not a supergroup, the vector 
index $a$ and chiral spinor index\foot*{We use upper and lower spinor
indices to distinguish the two chiralities, and 16$\ss\times$16 
$\ss\G$-matrices $\ss\G^a{}_{\a\b}$, $\ss\G^{a\,\a\b}$ (strictly speaking the 
$\ss\G$-matrices are 32$\ss\times$32 with these as off-diagonal blocks).} 
$\a$ never mix. Hence the components
$F_{ab},F_{a\b}$ and $F_{\a\b}$ can be treated as independent and, \eg,
 constraining a subset of them will not have any effect on the manifest 
supersymmetry.
Reading off the superspace torsion from the supersymmetry algebra we find that
superspace always has a non-trivial torsion component, namely $T_{\a\b}^{c}$.
In the case of flat superspace considered here,
$$
T_{\a\b}^{c}=2{\Gamma}_{\a\b}^{c}\eqn
$$
is the only non-zero one.

In this case, the component form of the superspace BI becomes 
$$\eqalign{
D_{(\a}F_{\b\g)}+2\G_{(\a\b}^cF_{|c|\g)}&=0\komma\cr
2D_{(\a}F_{\b)c}+D_{c}F_{\a\b}+2\G_{\a\b}^dF_{dc}&=0\komma\cr
D_{\a}F_{bc}+2D_{[b}F_{c]\a}&=0\komma\cr
D_{[a}F_{bc]}&=0\punkt\cr}\Eqn\allBI
$$

As is well-known the lowest order ten-dimensional supersymmetric 
Yang--Mills theory is obtained
by choosing the constraint [\NilssonSYM]
$$
F_{\a\b}=0\punkt\Eqn\Falphabetaiszero
$$
The vector part of this constraint, $(\G^a)^{\a\b}F_{\a\b}=0$, is
a so called conventional constraint [\ConvConstr] 
which must be imposed in order to
eliminate an unwanted extra vector potential appearing at the first 
$\theta$ level in $A_{\a}$. However, this can always be done without 
affecting the
supersymmetry since it just amounts to a shift of the superfield 
$A_{\a}$ by the
vector part of $F_{\a\b}$. 
To see how the equations of motion emerge, we 
insert the constraint (\Falphabetaiszero) into the above 
component equations (\allBI) leading to the 
following set of equations:
$$\eqalign{
\G_{(\a\b}^cF_{|c|\g)}&=0 \komma\cr
D_{(\a}F_{\b)c}+\G_{\a\b}^dF_{|d|c}&=0 \komma\cr
D_{\a}F_{bc}+2D_{[b}F_{c]\a}&=0 \komma\cr
D_{[a}F_{bc]}&=0 \punkt\cr}\Eqn\allBIonshell
$$
Instead of just presenting the solution to these equations, we will 
here take the opportunity
to elaborate, in a rather simple situation, on the different steps 
needed to obtain it. The same procedure will be followed also
in the next section where the equations as well as the 
process of finding the solution is 
significantly more complicated.

It is convenient to first analyze in detail the representation content of the 
equations as well as of the fields. Then the goal is to
derive all the 
algebraic relations between the irreducible fields that are hidden in 
these equations.  Decomposing into irreducible representations the symmetric 
product of three spinor representations we find\foot*{\epsfysize=24pt
We denote irreducible
representations of the Lorentz group by highest weight Dynkin labels
according to the standard enumeration
\vskip1pt\epsffile{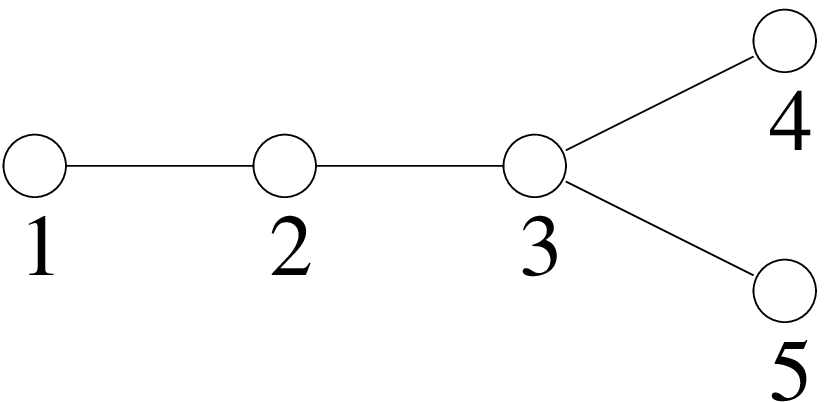}} 
$\otimes^3_s(00010)=(00030)\oplus(10010)$.
Thus the first Bianchi identity above can only produce restrictions on fields 
transforming in the two representations (00030) and (10010). Decomposing
also $F_{a\b}$ we find $(00001)\oplus(10010)$ which we write as
$$
F_{a\b}=\tF_{a\b}+\G_{a\,\b\g}\lambda^{\g}\komma\eqn
$$
where $\tF_{a\b}$ is $\G$-traceless, $\G^{a\,\a\b}\tF_{a\b}=0$. 
$\l$ will turn out to be the physical 
on-shell 
spinor field. Comparing the equation content to the field content we see 
immediately
that no information can be obtained about the spinor $\l$ while there is one 
equation for
the field in representation $(10010)$. 
In fact one finds that this equation puts
this field to zero. Thus the $(\a\b\g)$ Bianchi identity 
reduces to
$$
F_{a\b}=\G_{a\,\b\g}\lambda^{\g}\punkt\Eqn\DimThreeHalvesF
$$
Turning to the second component Bianchi identity 
(with index structure $a(\b\g)\,$) 
we repeat the above steps and obtain the table:
$$
\matrix{
D_\b\l^\c\hfill&:\hfill
&(00000)&\oplus&(00011)&\oplus&(01000)&&&&\cr
F_{ba}\hfill&:\hfill
&&&&&(01000)&&&&\cr
\hbox{BI}\hfill&:\hfill
&(00000)&\oplus&(00011)&\oplus&(01000)&\oplus&(10020)&\oplus&(20000)\cr
}\eqn
$$
Here the explicit form of the field content at first order in the $\theta$ 
expansion of $\l$
has been written in the equivalent form of a 
covariant derivative on $\l^{\a}$ as
$$
D_\a\l^\b=\d_\a{}^\b\L+\half\G^{ab}{}_\a{}^\b\L_{ab}+
\fr{4!}\G^{abcd}{}_\a{}^\b\L_{abcd}
\punkt\Eqn\DLambdaDecomposition
$$
The BI yields 
$$
\L=0\komma\qquad\L_{ab}=F_{ab}\komma\qquad\L_{abcd}=0\punkt\Eqn\LambdaSols
$$

The third BI, of dimension ${5\over2}$, reads
$$
D_\a F_{ab}=2(\G_{[a}D_{b]}\l)_\a\komma\eqn
$$
which when inserted into eq. (\DLambdaDecomposition) together with the 
solution for the $\L$'s (\LambdaSols) 
turns into $D_\a D_\b\l^\g=\G^{ab}{}_\b{}^\g(\G_aD_b\l)_\a$.
Contraction of this equation with $\G_c^{\a\b}\G^c_{\g\d}$ gives
the equation of motion for the spinor,
$$
\Dslash\l=0\punkt\eqn
$$
Acting on this equation with a spinor derivative and using eq.
(\DLambdaDecomposition) gives the equation of motion for the vector,
$$
D^bF_{ab}-\l\G_a\l=0\komma\eqn
$$
where the second term arises from pulling the spinor derivative through
the vector derivative and using the dimension-${3\over2}$ field
strength (\DimThreeHalvesF).

\section\solving{Solving the superspace Bianchi identities off-shell}In this 
section we relax the constraint $F_{\a\b}=0$ and 
derive what we will refer to as the off-shell equations. These equations
will include dynamical equations for $A_a$ and $\l_{\a}$ which however contain
other unspecified auxiliary fields taking the theory off the mass-shell 
derived from $F_{\a\b}=0$. The effects of relaxing the constraint were
discussed in ref. [\NilssonSYM], which ruled out the possibility of
constructing an off-shell lagrangian.

As discussed in the previous section the constraint $F_{\a\b}=0$ on
the dimension-1 field strength
puts the theory on the ordinary (lowest order) mass shell.
In order to relax it, we set [\NilssonSYM]
$$
F_{\a\b}=\fr{5!}\G_{\a\b}^{a_1\ldots a_5}J_{a_1\ldots a_5}\komma\Eqn\dimoneF
$$
where we choose  $J$ to be anti-selfdual, \ie, $J\in(00020)$.
In principle, there could be a $\G^{(1)}$ term, which is set to zero
by a conventional constraint as explained above.
At dimension $\Fr32$, $F$ is expanded as
$$
F_{a\a}=\tF_{a\a}+(\G_a\l)_\a\komma\eqn
$$
where $\tF\in(10010)$ is $\G$-traceless, and $\l\in(00001)$ 
($\l$ is the physical spinor field).
At the same time, $D_\a J_{abcde}$ is expanded according to 
$(00010)\otimes(00020)=(00030)\oplus(00110)\oplus(10010)$:
$$
DJ_{abcde}=\tJ_{abcde}+10\G_{[ab}\tJ_{cde]}+5\G_{[abcd}\tJ_{e]}\komma\eqn
$$
whose different irreducible parts can be reexpressed in terms of covariant derivatives on $J_{abcde}$ by means of the following inversion formul\ae:
$$
\eqalign{
\tJ_a&=\fr{1680}\G^{bcde}DJ_{bcdea}\komma\cr
\tJ_{abc}&=-\fr{12}\G^{de}DJ_{deabc}-\fr{224}\G_{[ab}\G^{defg}DJ_{|defg|c]}
	\komma\cr
\tJ_{abcde}&=DJ_{abcde}+\Fr56\G_{[ab}\G^{fg}DJ_{|fg|cde]}
	+\fr{24}\G_{[abcd}\G^{fghi}DJ_{|fghi|e]}\punkt\cr
}\Eqn\TildeJs
$$
The first equation to be solved is the one in (\allBI) of 
dimension $\Fr32$. It reads
$$
0=D_{(\a}F_{\b\c)}+T_{(\a\b}{}^aF_{|a|\c)}\komma\Eqn\ThreeHalvesBI
$$
and contains the irreps $\otimes^3_s(00010)=(00030)\oplus(10010)$.
Analyzing  eq. (\ThreeHalvesBI) we find that it sets $\tJ_{abcde}$ to zero,
leaves $\tJ_{abc}$ and $\l$ unconstrained and relates $\tJ_a$ and 
$\tF_{a}$ in the following way:
$$
\tF_{a\a}=-7\tJ_{a\a}\punkt\eqn
$$
The vanishing of the (00030) component is absolutely essential, and
has a cohomological interpretation (see the following section).
It is the only condition that the superfield $J_{abcde}$ has to satisfy.
Once it is fulfilled, the modified equations of motion follow.

At dimension 2 the relevant equation reads
$$
\eqalign{
0&=D_aF_{\a\b}-2D_{(\a}F_{|a|\b)}+T_{\a\b}{}^bF_{ba}\cr
&=\fr{5!}\G_{\a\b}^{a_1\ldots a_5}D_aJ_{a_1\ldots a_5}
	+14D_{(\a}\tJ_{|a|\b)}-2\G_{a(\a|\c|}D_{\b)}\l^\c
	+2\G_{\a\b}{}^bF_{ba}\punkt\cr
}\eqn
$$
The irreducible content of the various quantities is
$$
\matrix{
D_aJ_{a_1\ldots a_5}\hfill&:\hfill
&&&(00011)&&&\oplus&(10020)&&\cr
D_{(\a}\tJ_{|a|\b)}\hfill&:\hfill
&&&(00011)&\oplus&(01000)&\oplus&(10020)&&\cr
D_\b\l^\c\hfill&:\hfill
&(00000)&\oplus&(00011)&\oplus&(01000)&&&&\cr
F_{ba}\hfill&:\hfill
&&&&&(01000)&&&&\cr
\hbox{BI}\hfill&:\hfill
&(00000)&\oplus&(00011)&\oplus&(01000)&\oplus&(10020)&\oplus&(20000)\cr
}\eqn
$$
(in the second row, symmetrisation $(\a\b)$ has been used, as well as
the property $\tJ\sim DJ$, which takes away $(20000)$).
Schematically, the equations are
$$
\matrix{
(00000)&:&D_\a\l^\a=0\komma\hfill\cr
(00011)&:&D^2J+(D\G^{(4)}\l)\sim0\komma\hfill\cr
(01000)&:&D^2J+(D\G^{(2)}\l)+F\sim0\komma\hfill\cr
(10020)&:&D^2J\sim0\punkt\hfill\cr
}\eqn
$$
At the second level in $J$, one has
the irreps $(00011)\oplus(01000)\oplus(10020)\oplus(00120)
\oplus(01011)\oplus(10100)$, 
of which only the first three take part in the BI.
We write
$$
\eqalign{
D_{[\a}D_{\b]}J_{abcde}&=10\G_{[abc}K_{de]}+\half\G_{abcde}{}^{fg}K_{fg}\cr
	&\qquad+10\G_{[ab}{}^fK_{|f|cde]}+\Fr56\G_{[abcd}{}^{fgh}K_{|fgh|e]}\cr
	&\qquad+\Fr52(\G_{[a}{}^{gh})_{\a\b}S_{\vert gh\vert bcd,e]}
	+\ldots\komma\cr
D_{(\a}D_{\b)}J_{abcde}
	&=-\fr2T_{\a\b}{}^fD_fJ_{abcde}-\fr2[F_{\a\b},J_{abcde}]\cr
	&=-\G_{\a\b}^fD_fJ_{abcde}
	-\fr{2\cdot5!}\G^{fghij}_{\a\b}[J_{fghij},J_{abcde}]\cr
}\eqn
$$
(the $[J,J]$ part only contains the representations 
$(00120)\oplus(10100)$ which do not enter the dim 2 BI).
The inversions for the $K$'s will be used later and read
$$
\eqalign{
K_{ab}&=\fr{5376}(D\G^{cde}D)J_{cdeab}\komma\cr
K_{abcd}&=\fr{480}(D\G_{[a}{}^{fg}D)J_{|fg|bcd]}\punkt\cr
}\Eqn\KintermsofJ
$$
We also decompose $D\l$ as
$$
D_\a\l^\b=\d_\a{}^\b\L+\half\G^{ab}{}_\a{}^\b\L_{ab}+\fr{4!}\G^{abcd}{}_\a{}^\b\L_{abcd}
\punkt\eqn
$$
The equations become:
$$
\eqalign{
(00000):\quad0&=\L\komma\cr
(00011):\quad0&=\Fr7{30}D^fJ_{fabcd}+2K_{abcd}-\L_{abcd}\komma\cr
(01000):\quad0&=\Fr{28}5K_{ab}+\L_{ab}-F_{ab}\komma\cr
(10020):\quad0&=S_{abcde,f}+\Fr56\left(D_fJ_{abcde}+D_{[a}J_{bcde]f}
	-\eta_{f[a}D^gJ_{|g|bcde]}\right)\punkt\cr
}\Eqn\DimThreeHalvesEq
$$

There is one important consistency check here: when $\tJ_{abcde}$ in the
representation (00030) vanishes at dimension ${3\over2}$, there is no
new independent (10020) at dimension 2 (see figure 1). 
The equation obtained from
acting with one spinor derivative on $\tJ_{abcde}=0$ will contain 
a part in (10020) that had better coincide with the one in 
eq. (\DimThreeHalvesEq), if we are not to get a differential constraint
on $J$. We have checked that this is the case.
Similar consistency conditions arise at higher dimensions, and are
always automatically fulfilled once the (00030) representation at dimension
$\Fr32$ vanishes.
The (00030) constraint also implies that the (00120) part of
$D_{[\a}D_{\b]}J_{abcde}$ is expressible in terms of the one in $[J,J]$. 
The $(10100)\oplus(01011)$ part remains unconstrained.

The dimension-$\Fr52$ Bianchi identity reads
$$
0=2D_{[a}F_{b]\a}+D_\a F_{ab}=-14D_{[a}\tJ_{b]\a}-2\G_{[a|\a\b|}D_{b]}\l^\b
	+D_\a F_{ab}\punkt\eqn
$$
Inserting this expression for $D_\a F_{ab}$ into one spinor derivative
on $D_\a\l^\b$ from above gives
$$
\eqalign{
D_\a D_\b\l^\g
&=\half\G^{ab}{}_\b{}^\g\bigl[2(\G_a D_b\l)_\a+14D_a\tJ_{b\a}
			-\Fr{28}5D_\a K_{ab}\bigr]\cr
&\qquad+\fr{24}\G^{abcd}{}_\b{}^\g\bigl[\Fr7{30}D_\a D^fJ_{fabcd}
			+2D_\a K_{abcd}\bigr]\komma\cr
}\eqn
$$
and contracting with $\G_g{}^{\a\b}\G^g{}_{\d\g}$ gives
(a preliminary form of) the equation of motion for $\l$:
$$
0=\Dslash\l+\Fr35D^a\tJ_a+\fr{3600}\G^{abcd}DD^fJ_{fabcd}
	+\Fr6{25}\G^{ab}DK_{ab}+\fr{420}\G^{abcd}DK_{abcd}\punkt
\Eqn\PrelLambdaEOM
$$

In order to simplify the equation of motion (\PrelLambdaEOM) for $\l$
we need to expand the last three terms.
As seen in figure 1, there is one spinor (00010) at dimension $\Fr52$ in $J$.
We parametrise it as
$$
D_{[\a}D_\b D_{\g]}J^{abcde}
=30\G^{[abc}_{[\a\b}(\G^{de]}_{\phantom{[\a]}}\psi)^{\phantom{[a]}}_{\g]}
+\ldots\komma\Eqn\PsiinJ
$$
with the inversion
$$
\psi_\a=-\fr{840\cdot3!\cdot5!}		
	\G_{abc}{}^{\b\g}\G_{de\,\a}{}^\d D_{[\b}D_\g D_{\d]}
	J^{abcde}\punkt\eqn
$$
Since there is no (00010) in $\wedge^3(00010)\otimes(00002)$, the right
hand side of eq. (\PsiinJ) is automatically anti-selfdual.
We insert this into one spinor derivative on eq. (\KintermsofJ) to
obtain, after some calculation, 
$$
\eqalign{
\G^{ab}DK_{ab}&=-\Fr{225}2\psi+\Fr52D^a\tJ_a
	+\fr{2016}\G^{abcde}[\l,J_{abcde}]\komma\cr
\G^{abcd}DK_{abcd}&=-1260\psi-140D^a\tJ_a
	-\fr{36}\G^{abcde}[\l,J_{abcde}]\punkt\cr
}\eqn
$$
The $\psi$ terms come from the totally antisymmetrised product of three
derivatives, while terms with mixed symmetrisation give the  $\tJ$ 
and commutator terms from torsion and curvature, respectively. We use
$D_\d D_{[\b}D_{\g]}=D_{[\d}D_\b D_{\g]}+\Fr43D_{(\d}D_{\b)}D_\g
-\Fr23D_\b D_{(\d}D_{\g)}$, where antisymmetrisation $[\b\g]$ is understood
in the right hand side.
The equation of motion for $\l$ is 
$$
0=\Dslash\l-30\psi+\Fr43D^a\tJ_a
	+\Fr5{126\cdot5!}	
	\G^{abcde}[\l,J_{abcde}]\punkt\Eqn\LambdaEOM
$$

To derive the equation of motion for $F$, we start from 
the equation of motion for $\l$, act with a spinor derivative
and contract with a $\G$ matrix, \ie, write eq. (\LambdaEOM)
as $0=\L^{(1)}+\L^{(2)}+\L^{(3)}+\L^{(4)}\equiv\L$ and consider the equation 
$0=D\G_a\L\equiv L_a\equiv L^{(1)}_a+L^{(2)}_a+L^{(3)}_a+L^{(4)}_a$.
Then,
$$
\eqalign{
L^{(1)}_a&=16D^b(F_{ab}-\Fr{28}5K_{ab})-14\{\l,\tJ_a\}-16\l\G_a\l\komma\cr
L^{(2)}_a&=576w_a-\Fr{64}3D^bK_{ab}-\Fr{40}9\{\l,\tJ_a\}
	-\Fr{40}3\tJ_b\G_a\tJ^b+\Fr2{105}\tJ_{bcd}\G_a\tJ^{bcd}\cr
	&\qquad+\Fr{52}{189}[K_{bcde},J_a{}^{bcde}]
	+\Fr2{567}[D^fJ_{fbcde},J_a{}^{bcde}]\komma\cr
L^{(3)}_a&=-\Fr{256}{15}D^bK_{ab}+\Fr83\{\l,\tJ_a\}
	-\Fr{56}3\tJ_b\G_a\tJ^b\komma\cr
L^{(4)}_a&=-\Fr{50}9\{\l,\tJ_a\}+\Fr{20}{189}[K_{bcde},J_a{}^{bcde}]
	+\fr{81}[D^fJ_{fbcde},J_a{}^{bcde}]\punkt\cr
}\eqn
$$
Here, we have defined the vector $w$ at fourth level in $J$ by
$$
D_{[\a}D_\b D_\g D_{\d]}J^{abcde}
	=60\G^{[abc}_{[\a\b}\G^{de]f}_{\g\d]}w^{\phantom{[\a]}}_f
	+\ldots\komma\eqn
$$
with the inversion
$$
w_a=\fr{4032\cdot4!\cdot5!}\G^{[\a\b}_{abc}\G^{\g\d]}_{def}
	D_\a D_\b D_\g D_\d J^{bcdef}\komma\eqn
$$
and used $D_\a D_{[\b}D_\g D_{\d]}=D_{[\a} D_\b D_\g D_{\d]}
+\Fr32D_{(\a}D_{\b)}D_\g D_\d-D_\b D_{(\a}D_{\g)}D_\d
+\fr2D_\b D_\g D_{(\a}D_{\d)}$ 
(antisymmetrisation $[\b\g\d]$ understood) in $L^{(2)}$.

The equation of motion for $A$ is thus
$$
\eqalign{
0&=D^bF_{ab}-\l\G_a\l-8D^bK_{ab}+36w_a-\Fr43\{\l,\tJ_a\}
-2\tJ_b\G_a\tJ^b+\fr{140\cdot3!}\tJ_{bcd}\G_a\tJ^{bcd}\cr
&\qquad+\fr{42}[K_{bcde},J_a{}^{bcde}]
	+\fr{42\cdot4!}[D^fJ_{fbcde},J_a{}^{bcde}]\punkt
}\eqn
$$
\vskip3\parskip
\epsfxsize=.95\hsize
\epsffile{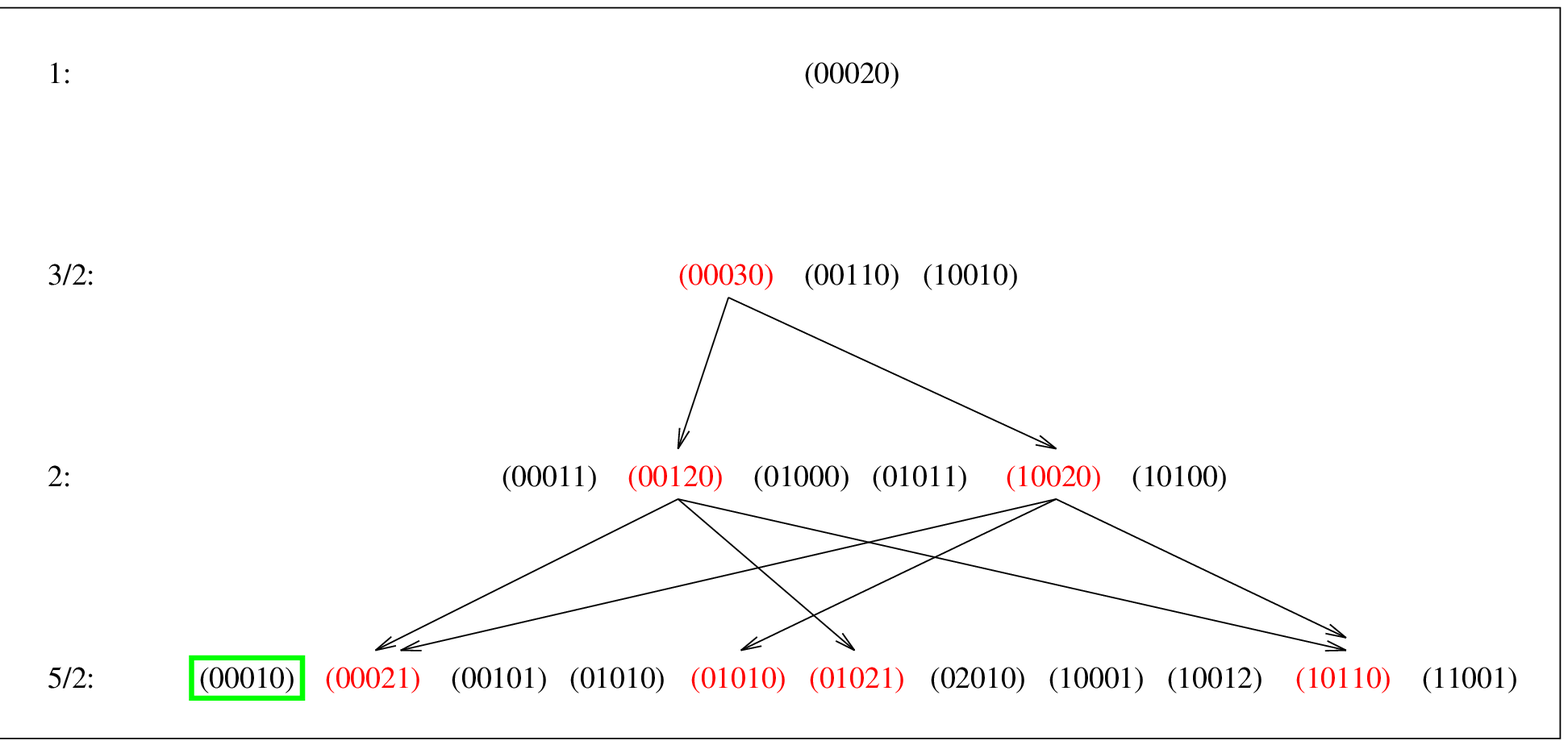}
{\it Figure 1. The representations in $J$ up to dimension $\Fr52$. 
The arrows show how the
$(00030)$ constraint propagates. The boxed representation is the spinor
$\psi$
responsible for the right hand side of the equation of motion for
$\l$. The vector $w$ at dimension 3, occurring in the equation of motion 
for $A$, is also outside the (00030) superfield.}
\vskip2\parskip

\section\Cohomology{Fields and interactions from spinorial cohomology}In
this section, we would like to present an argument that gives conceptual
support to the procedure we have adopted for constructing higher-order
corrections to the ordinary super-Yang--Mills theory.

The gauge potentials are $A_\a$ and $A_a$. However, the spinor potential
already contains a vector (of correct dimension) at the $\theta$ level,
and this is the reason why a conventional constraint is needed in order
to have {\it one} vector potential. This constraint is 
$$
\G_a^{\a\b}F_{\a\b}=0\komma\eqn
$$
which implies that (in the abelian case, for simplicity) 
$$
A_a=-\fr{32}D\G_aA\punkt\eqn
$$
The rest of $F_{\a\b}$, which lies in (00020), does not contain $A_a$.
We have also noted that part of the dimension-$\Fr32$ Bianchi identity
states the vanishing of the (00030) component of $D_\a F_{\b\g}$.

These observations make it natural to consider, 
not the sequence of completely symmetric
representations in spinor indices, but a restriction of it, namely 
the sequence of Spin(1,9) representations
$$
(00000)\Dlra0(00010)\Dlra1(00020)\Dlra2
\ldots\Dlra{n-1}(000n0)\Dlra{n}\ldots\Eqn\Complex
$$
The representation $r_n\equiv(000n0)$ is the part of the totally symmetric
product of $n$ chiral spinors that has vanishing $\G$-trace, and may
be represented tensorially as 
$C_{\a_1\ldots\a_n}=C_{(\a_1\ldots\a_n)}$, 
$\G_a{}^{\a_1\a_2}C_{\a_1\a_2\a_3\ldots\a_n}=0$.
For $n=2$, $C$ is an anti-selfdual five-form, for $n=3$ a $\G$-traceless
anti-selfdual five-form spinor, etc.

The operator $\D_n$: $r_n\lra r_{n+1}$ can schematically be written
as $\D_nC_n=\Pi(r_{n+1})DC_n$, where $D$ is the exterior covariant derivative
$D=d\theta^\a D_\a$ and $\Pi(r_n)$ is the algebraic projection
from $\otimes^n_s(00010)$ to $(000n0)$. It is straightforward to write
an explicit tensorial form for $\D$ by subtracting $\G$-traces from $DC$,
but it will not be used here.

It is straightforward to show that, for an abelian gauge group
and standard flat superspace,
the sequence (\Complex) forms
a complex, \ie, that $\D^2=0$. This follows simply from the fact that
while $\{D_\a,D_\b\}=-T_{\a\b}{}^cD_c$, the torsion only has a component
$2\G_{\a\b}{}^c$ which is projected out by $\Pi(r_n)$.

The question immediately arises whether this complex contains any
non-trivial cohomology. It is sometimes stated that fermionic
cohomology is trivial. This is true when one considers the complex
of symmetric multi-spinors but, as we will see, 
not when projected on the $r_n$'s.

To investigate the content at each $n$, one has to expand the superfields
$C_{\a_1\ldots\a_n}$ in irreducible component fields. 
The representation occurring at level $\ell$ (multiplying $\theta^\ell$) 
in $C_n$ is given
as $r_n^\ell\equiv\wedge^\ell(00010)\otimes(000n0)$. 
We would like to keep track of
dimensions of fields, so we give $C_n$ mass dimension ${n\over2}$, so that
a component of $C_n$ at level $\ell$ has dimension ${n+\ell\over2}$.
Likewise, the cohomology is decomposed as 
$\H^n\equiv\hbox{Ker}\D_n/\hbox{Im}\D_{n-1}=\oplus_\ell\H^{n,\ell}$.
The cohomology is purely algebraic, and is calculated simply
as $\H^{n,\ell}=r_n^\ell\ominus r_{n-1}^{\ell+1}\ominus r_{n+1}^{\ell-1}$
where the second and third terms represent the level $\ell$
contents of Im$\D_{n-1}$ and (Ker$\D_{n}$)${}^\perp$
respectively, and where ``$\ominus$'' means subtraction of an
irreducible representation only if it is already present.

To be explicit, we present the calculation of $\H^1$. We have 
$$
\eqalign{
\H^{1,0}&=r_1^0\ominus r_0^1=(00010)\ominus(00010)=0\komma\cr
\H^{1,1}&=r_1^1\ominus r_0^2\ominus r_2^0\cr
	&=(00010)\otimes(00010)\ominus\wedge^2(00010)\ominus(00020)\cr
	&=(10000)\oplus(00100)\oplus(00020)\ominus(00100)\ominus(00020)\cr
	&=(10000)\komma\cr
\H^{1,2}&=r_1^2\ominus r_0^3\ominus r_2^1\cr
	&=\wedge^2(00010)\otimes(00010)\ominus\wedge^3(00010)
		\ominus(00010)\otimes(00020)\cr
	&=(00110)\oplus(01001)\oplus(10010)\oplus(00001)\cr
	&\qquad\ominus(01001)\ominus[(00030)\oplus(00110)\oplus(10010)]\cr
	&=(00001)\komma\cr
\H^{1,3}&=r_1^3\ominus r_0^4\ominus r_2^2\cr
	&=\wedge^3(00010)\otimes(00010)\ominus\wedge^4(00010)
		\ominus\wedge^2(00010)\otimes(00020)\cr
	&=(00011)\oplus(01000)\oplus(01011)\oplus(02000)\oplus(10002)
		\oplus(10100)\cr
	&\qquad\ominus[(02000)\oplus(10002)]\cr
	&\qquad\ominus[(00011)\oplus(00120)\oplus(01000)\oplus(01011)
		\oplus(10020)\oplus(10100)]\cr
	&=0\punkt\cr
}\eqn
$$
Higher $\H^{1,\ell}$ vanish. Using the dimensions instead of the level,
we find that
$$
\H^1=(10000)_1\oplus(00001)_{3\over2}\punkt\eqn
$$
The interpretation of this result is clear. $C_1$ is the spinor
potential $A_\a$, and the cohomology represents
the physical component fields in the vector multiplet, 
$A_a$ of dimension 1 and 
$\lambda^\a$ of dimensions $\Fr32$. Subtraction of Im$\D_0$ means counting
fields modulo gauge 
transformations, and subtraction of (Ker$\D_1$)${}^\perp$ 
reflects the constraint
$F_{\a\b}=0$. The cohomology is not supersymmetric, since the
manipulations so far only involved spinorial derivatives. Imposing the
complete Bianchi identities, as shown in section {\old3}, leads to the
equations of motion for the component fields.

Let us continue with the second cohomology $\H^2$. The calculation
is completely analogous, and the result is that it contains a
spinor (00010) at $\ell=3$ and a vector at $\ell=4$:
$$
\H^2=(00010)_{5\over2}\oplus(10000)_3\punkt\eqn
$$
These components match the ones of the currents occurring in the
equations of motion for $\l$ and $A$. We can thus identify a deformation
of the theory with a field strength $F_{\a\b}$ being an element of $\H^2$. 
Now, subtraction of Im$\D_1$ means counting modulo field redefinitions,
and subtraction of (Ker$\D_2$)${}^\perp$ is related to the 
dimension-$\Fr32$ Bianchi
identity, which implies the vanishing of the (00030) part of $D_\a F_{\b\g}$.

For a non-abelian gauge group, one can imagine starting from the
undeformed theory with $F_{\a\b}=0$, and trying to deform it 
infinitesimally by introducing some non-zero $F_{\a\b}$.
Then $\D$ is defined with respect to the undeformed theory, and an
infinitesimal deformation is an element of the cohomology. Finite
deformations demand a more refined analysis taking into account the 
interplay between terms of different orders in an expansion parameter
(\eg\ $\a'$).
Field strengths expressible as $F=\D a$ can be absorbed into the spinor
potential through a field redefinition (this will be used explicitly
in section {\old5} and in a forthcoming publication [\CederwallSYMFFOUR]).
We then get information about physically inequivalent deformations
of supersymmetric gauge theories.

The statement that (part of) the field strength belongs to a 
non-trivial cohomology class na\"\i vely seems to contradict the statement that
it is obtained from a gauge potential (which we made explicit use of
when deriving the field equations from the Bianchi identities). 
However, when the rest of the
Bianchi identities are imposed, they will imply equations of motion. 
Let us take the
spinor equation as example. It reads $\Dslash\l\sim\psi+\ldots$. If
the cohomology class is trivial, the right hand side will be expressible
as $\Dslash\mu$, so that the deformation is removed by a field redefinition.
If the cohomology class is non-trivial, on the other hand, the equation
of motion states that the right hand side is  $\Dslash\l$, and in this
sense the equations of motion resolve the cohomology.

It is instructive to consider the analogous complex for six-dimensional
super-Yang--Mills with $N=(1,0)$ supersymmetry [\NilssonSixDSYM]. 
This theory has an
off-shell formulation in terms of the vector, the spinor and a triplet
of auxiliary scalars of dimension 2. 
The complex is 
$$
(000)(0)\Dlra0(100)(1)\Dlra1(200)(2)\Dlra2
\ldots\Dlra{n-1}(n00)(n)\Dlra{n}\ldots\Eqn\Complex
$$
Indeed, the first cohomology is
$$
\H^1=(010)(0)_1\oplus(001)(1)_{3\over2}\oplus(000)(2)_2\komma\eqn
$$
where the representations are given as standard Dynkin labels for
Spin(1,5)$\times$SU(2) (the second factor being the R-symmetry group). 
The second cohomology is trivial, which also
is expected---setting $F_{\a\b}$
to zero does not put the theory on-shell, and the value of $F_{\a\b}$
does not contain any information about interactions, it can be set to
zero by a field redefinition.

Finally, we would like to comment on a potentially interesting observation
concerning the ten-dimensional theory. While $\H^n$, $n\geq4$ seem to vanish,
the third cohomology $\H^3$ is non-trivial, and contains only a scalar 
of dimension 4, which is the dimension of a lagrangian. 
We do not yet understand what this signifies.

\section\conclusions{Conclusions and outlook}In this paper we 
use superspace techniques to 
derive a set of algebraic constraints on the 
irreducible field components in the superfield strengths $F_{AB}$.
Since some of these field components will correspond to composite operators
in theories with higher-order corrections, the algebraic constraints 
will provide restrictions on the possible explicit form of 
the corrections. In particular, the abelian Born--Infeld action must 
satisfy these restrictions as must its  non-abelian kin whose lagrangian
we have very limited information about.

Whereas $\kappa$-symmetry and non-linear supersymmetries are known to play
a very important role in dictating the form of the abelian Born--Infeld action,
it is not known how much of the non-linear structure
can be deduced from the maximal linear supersymmetry alone. We will return to
this question 
in a forthcoming publication [\CederwallSYMFFOUR], and conclude this section
by  briefly discussing the role of $F_{\a\b}$ in this context.

We introduced the five-form $J_{abcde}$ in eq. (\dimoneF). 
As follows from the results in section {\old3}, expressing $J_{abcde}$ 
in terms of the physical fields $F_{ab}$ and $\l_{\a}$ will give rise to
field equations with higher-order corrections, as long as the (00030)
constraint is fulfilled, \ie, $D_\a J_{abcde}\vert_{(00030)}$ vanishes.

We first observe that there are no corrections at order $\a'$. For
dimensional reasons, $F_{\a\b}$ has to be proportional to $\l^2$, which
does not contain the representation (00020).
Then, starting at order $\a'^2$, there are two types of possible terms, modulo
the lowest order field equations ($A,B,\ldots$ are adjoint 
gauge group indices, not to be confused with $A=(a,\a)$ used earlier): 
$$
\eqalign{
J^A_{abcde}&=\half\a'^2 M^A{}_{BCD}(\l^B\G^f\G_{abcde}\G^g\l^C)F^D_{fg}\cr
&+\fr6\a'^2 N^A{}_{BC}\left(D_{[a}\l^B\G_{bcd}D_{e]}\l^C-\hbox{ dual}\,\right)
\punkt\cr
}\Eqn\JExpansion
$$
These satisfy the (00030) constraint at linear order, which is
easily seen by acting with a spinor derivative and perform 
tensor multiplication of the representations of the fields.
So far, $M$ and $N$ are kept arbitrary, but with the manifest symmetry $(BC)$. 
To derive the equations of motion we need some of the higher components of $J$.
This will be done in a following paper [\CederwallSYMFFOUR], where the
complete action at order $\a'{}^2$ will be constructed.
For the moment we will content ourselves with extracting the physically
relevant deformations out of $J$. We want to determine which types of
terms can be removed by fields redefinitions and which can not.
The field redefinitions are taken care of by shifting $A_\a$ as explained
in the previous section. These shifts can be of three independent forms
at this order in $\a'$, namely
$$
\eqalign{
\d A^A_\a&=\fr6\a'^2m^A{}_{BCD}(\l^B\G^{abc}\l^C)(\G_{abc}\l^D)_\a\cr
	&+\a'^2n^A{}_{BCD}(\l^B\G^a\l^C)(\G_a\l^D)_\a\cr
	&+\a'^2p^A{}_{BC}F^{B\,ab}(\G_aD_b\l^C)_\a\punkt\cr
}\Eqn\FieldRedefinitions
$$
When we calculate $\d F_{\a\b}=(\D\d A)_{\a\b}$, 
only the first and third of these contribute to (00020) (any contribution
to the vector part is removed by an accompanying redefinition of $A_a$,
so that the conventional constraint remains unaffected).
The third one is used to get rid of any ``$D\l D\l$'' terms in eq.
(\JExpansion), so these can be discarded as irrelevant.
Examining the first (``$\l\l\l$'') term in (\FieldRedefinitions), 
we observe that it has mixed symmetry
in the $\l$'s (\ie, neither the completely symmetric nor the
completely antisymmetric product of three (00001)'s contain (00010)).
Consequently, $\d J_{abcde}$ is proportional to
$\a'{}^2m^A{}_{B[CD]}(\l^B\G^f\G_{abcde}\G^g\l^C)F^D_{fg}$.
Since the combination of fields contracting $M$ and $m$ are manifestly
symmetric in $(BC)$, this field redefinition is used to remove the part
of $M$ with mixed symmetry, and the remaining relevant part of $M$ is
only the one completely symmetric in $(BCD)$. For any gauge group, this
seems to imply that $M$ is symmetric in all four indices.
If the gauge group is SU($N$), there are two available tensors,
$d_{(AB}{}^Ed_{CD)E}$ and $\d_{(AB}\d_{CD)}$. The symmetric trace
is one specific combination of these.
 
This discussion shows that any $F^4$ terms in the action (that cannot be
removed by field redefinitions) are completely symmetric in the
adjoint indices. This does not mean that all terms at this order in
$\a'$ are contracted with symmetric tensors; we have already seen that
the field equations contain commutator terms, and we expect terms
like ``$F\l^4$'' contracted with an $M$ and a structure constant $f$.

It will be very interesting to examine to which degree linear supersymmetry
alone determines the higher-order corrections. Clearly, once the first
correction is postulated, higher ones are necessary. In our framework,
we see this as the need for cancellation of the (00030) constraint
at order $\a'{}^4$. It is yet an open question to what degree there
remains an arbitrariness in this procedure, \ie, if new (non-polynomial)
invariants arise that start at order $\a'{}^4$ and higher.
Our intuition leads us to suspect that the higher-order interactions
are not unique (we have at least evidence that an independent invariant
exists that starts at order $\a'{}^3$ [\Kitazawa]), 
and that a second, non-linearly realised, supersymmetry
is necessary in order to determine the Born--Infeld action.
We envisage that the techniques of the present paper, where all 
the dynamics is encoded in the relatively simple object $F_{\a\b}$,
and where the (00030) constraint is the only condition needed to
consider in the iterative procedure,
is suited for addressing such questions.

We would like to compare the results from the present approach with 
those obtained by Bergshoeff \etal\ [\BergshoeffKAPPA] 
by demanding a non-abelian $\k$-symmetry
(with a spinor parameter in the adjoint of the gauge group).
In that paper, quartic terms were found at order $\a'{}^2$ that are not
contracted by a completely symmetric tensor $M$. Our results seem to
indicate that such terms, in any supersymmetric gauge theory, are trivial
and removable by field redefinitions. 

The issues and techniques discussed in this paper can be 
directly taken over to
eleven-dimensional supergravity and the higher-order corrections generated by
string/M-theory. In fact, one of the reasons for the study conducted here
is to investigate the ideas in [\CGNN], where they were applied to M-theory, 
in the much simpler context of Yang--Mills theory. 


\acknowledgements
We would like to thank Gabriele Ferretti for discussions and for pointing
out references.
This work is partly  
supported by EU contract HPRN-CT-2000-00122 and by the Swedish
Natural Science Research Council. 
For some representation-theoretical considerations, such as tensor
products of representations, the program 
LiE [\LiE] has been very useful.

\vfill\eject
\null\vskip-48pt
\xrm\refout 

\end